\documentclass[twocolumn]{revtex4}

\usepackage{setspace} \usepackage{amsmath,amssymb,amsfonts,amsthm}
\usepackage{graphicx}

\newcommand{\be}{\begin{equation}}
\newcommand{\ee}{\end{equation}}
\newcommand{\beq}{\begin{eqnarray}}
\newcommand{\eeq}{\end{eqnarray}}

\def\lsim{\hbox{ \raise.35ex\rlap{$<$}\lower.6ex\hbox{$\sim$}\ }}
\def\gsim{\hbox{ \raise.35ex\rlap{$>$}\lower.6ex\hbox{$\sim$}\ }}

\begin{document}
\title{On the possibility of Dark Energy from corrections to the Wheeler-DeWitt equation} 
\author{William Nelson and Mairi Sakellariadou}
\affiliation{King's College London,
Department of Physics, Strand WC2R 2LS, London, U.K.}

\begin{abstract}
We present a method for approximating the effective consequence of
generic quantum gravity corrections to the Wheeler-DeWitt equation.
We show that in many cases these corrections can produce departures
from classical physics at large scales and that this behaviour can be
interpreted as additional matter components. This opens up the
possibility that dark energy (and possible dark matter) could be large
scale manifestations of quantum gravity corrections to classical
general relativity. As a specific example we examine the first order
corrections to the Wheeler-DeWitt equation arising from loop quantum
cosmology in the absence of lattice refinement and show how the ultimate
breakdown in large scale physics occurs.

\vspace{.2cm}
\noindent

\end{abstract}

\maketitle

%%%%%%%%%%%%%% Paper body %%%%%%%%%%%%%%%%%%%%%%%%%
%%%%%%%%%%%%%%%%%%%%%%%%%%%%%%%%%%%%%%%%%%%%%

Quantum cosmology can be studied within the context of mini-superspace
models, reducing the full quantum field theory to a quantum mechanical
system of finite degrees of freedom. Applying this to the evolution of
a three metric results in the Wheeler-DeWitt equation (WDW)
equation~\cite{D_eath}, which breaks down near the classical big-bang
singularity.  Loop Quantum Gravity (LQG), another approach to
canonical quantisation, uses triads and
connections rather than metrics and extrinsic curvatures as the
basic variables, with more success~\cite{Ashtekar:2003hd}, whilst
string and brane theories remove the singular behaviour by the extra
dynamical states that become available at small scales. Irrespective
of the full underlying theory, the WDW equation must be recovered as a
semi-classical approximation to the dynamical equations of the theory,
if classical general relativity is to be produced at large scales. We
can then ask whether Quantum Gravity (QG) corrections can have any
effects on large scale physics.

At first sight it would appear unlikely that this is possible. However,
the wave-function solutions to the WDW equation oscillate and so their
derivatives can become significant at large scales. If the QG corrections
depend on the derivatives of the wave-function, then it is possible that
they become significant at macroscopic scales. This is the case for Loop
Quantum Cosmology (LQC), which predicts that the evolution of the universe
is dictated by a difference, rather than a differential equation. If the
lattice size of this difference equation is fixed, as was assumed until
recently~\cite{Ashtekar:2006uz}, not all the oscillations of the
wave-function can be supported, leading to a breakdown in large scale
classical physics~\cite{Ashtekar:2006uz,Nelson:2007wj}.

We investigate the effects of a general class of corrections to the WDW equation,
considering all possible derivative terms and show that, at least
initially, these corrections mimic the behaviour of additional matter
components. This raises the exciting possibility that QG
corrections could, at late times, produce a cosmological constant like
effect. It is also possible that such corrections may produce additional
matter components such as dark matter, however this would only be true
for a  limited class of corrections. What is generally true is that
QG correction terms that dominate at small scale can also produce
significant large scale effects, which can be well approximated by
the classical behaviour of additional matter sources.

By considering the mini-superspace model of an homogeneous and isotropic
universe, the WDW equation reads
\be
\label{eq:WdW0} 
\frac{1}{a}\frac{{\rm d}}{{\rm d}a} \left[ \frac{1}{a}\frac{{\rm
d}}{{\rm d}a} \left[ a\psi\left(a\right)\right]\right] -\frac{9k}{
16\pi^2 l_{\rm Pl}^2} a\psi\left(a\right) +\frac{3}{2\pi l_{\rm
Pl}^2}{\cal H}_\phi\psi\left(a \right)=0~
\ee
where $a$ is the scale factor, $k=0,\pm1$ is the curvature, ${\cal
H}_\phi$ is the matter Hamiltonian; in our units
$\hbar=c=1$\footnote{Note that the factor ordering is not unique.  We
will be interested only in the scaling behaviour of $\psi\left(
a\right)$ at large scales where the factor ordering is of no
consequence.}.

In general, ${\cal H}_\phi$ will have several terms, with different
dependence on $a$ and $\phi$. Being only interested in the large scale
($a \gg l_{\rm Pl}$) behaviour of ${\cal H}_\phi$, we expect one term
to dominate and we approximate it by
$[3/(2\pi l_{\rm Pl}^2)]{\cal H}_\phi = \epsilon\left(\phi\right)
a^\delta.$
With this approximation the WDW equation can  be written in as
\be
\label{eq:WdW1}
\frac{1}{a}\frac{{\rm d}}{{\rm d}a} \left[ \frac{1}{a}\frac{{\rm
d}}{{\rm d}a} \left[ a\psi\left(a\right)\right]\right] -\frac{9k}{
16\pi^2 l_{\rm Pl}^2} a\psi\left(a\right) +\epsilon\left(\phi\right)
a^\delta \psi \left(a\right)=0~.
\ee  
We only consider the flat case, so that solutions to 
Eq.~(\ref{eq:WdW1}) are analytically tractable.
If this equation breaks down at small scales, due to some
QG effects, extra terms should become
significant on these scales. These new QG correction 
terms will be a function of $\psi\left( a\right)$ and its derivatives,
so the full underlying equation can be written as
\beq\label{eq:WdW2}
\frac{1}{a}\frac{{\rm d}}{{\rm d}a} \left[ \frac{1}{a}\frac{{\rm d}}{{\rm d}a}
\left[ a\psi\left(a\right)\right]\right]
+ \epsilon\left(\phi\right)a^\delta
 \psi\left( a\right)\nonumber\\
 + a_0 f\left( \psi\left( a\right), \partial_a \psi, 
\partial^2_a \psi \cdots \right) =0~,
\eeq
where $a_0$ is the scale at which the new terms become important.  At
large scales, $a \gg a_0 $, one might naively assume that the QG 
corrections can simply be ignored, however this assumes that
$\psi\left( a \right)$ and all its derivatives remain small at all
scales, which is easily shown not to always be consistent. This has
been explored within the context of 
LQC~\cite{Ashtekar:2006uz, Nelson:2007wj,Bojowald:2007ra}, in
which the exact form of the $a_0$ corrections are known.

\begin{table}
 \begin{tabular}{|c|c|c|c|}
 \hline & $\rho$ & ${\cal H}_\phi$ & $\delta$ \\ \hline Matter &
  $a^{-3} $ & $ a^0 $ & $0$ \\ Radiation & $a^{-4} $ & $a^{-1}$&$-1$
  \\ Vacuum energy & $a^0 $ & $a^3$ & $3$\\ \hline
 \end{tabular}
 \caption{\label{table} The scaling of the energy density $\rho$ and
the Hamiltonian ${\cal H}_\phi=a^3\rho$ for matter,
radiation and vacuum energy with respect to both $a$ and $\mu$. 
The parameter $\delta$ is also given. Note that the energy density and
hence ${\cal H}_{\phi}$ of vacuum energy are negative.}
\end{table}

\begin{figure}
 \begin{center}
  \includegraphics[scale=0.75]{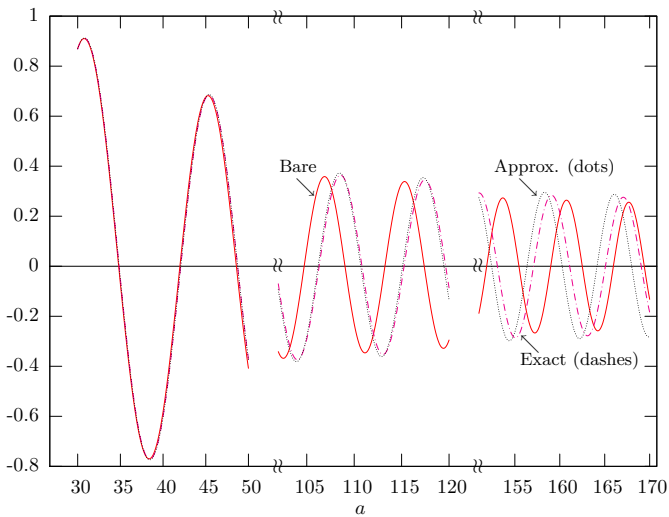}
  \caption{\label{fig1} Solutions to the WDW equation, for $\delta =
   0$ (i.e., matter dominated universe), $\epsilon(\phi)=5\times
   10^{-3}$, $a_0=9\times 10^{-6}$ (with $l_{\rm Pl}=1$). The QG
   corrections are $a_0a^2{\rm d}^2\psi \left(a\right)/{\rm d}^2
   a$. The solutions are calculated numerically: the {\it bare}
   solution (solid) excludes QG corrections entirely whilst the {\it
   approx.} solution (dots) approximates them as discussed in the text. For
   small $a$ the QG corrections do not play a role, however as $a$
   becomes larger the effects of the QG corrections become
   significant. The form of this deviation is well characterised by
   the addition of a second matter component, which is demonstrated by
   the accuracy of the approximate solution (see dotted line). For $
   a\approx 165$, the approximate correction term is $\approx 0.25
   \epsilon \psi\left(a\right)$ and hence can no longer be considered
   small compared to the matter component, $\epsilon \psi
   \left(a\right)$; the approximation breaks down.}
 \end{center}
\end{figure}

Assuming $a_0 \ll \epsilon\left( \phi\right)$, whilst
$f\left( \psi\left( a\right),\partial_a \psi,\cdots\right)$ is small,
we can find approximate solutions to Eq.~(\ref{eq:WdW2}) as
\beq\label{eq:sol}
 \psi\left(a\right) &\approx& C_1 \mbox{ \large{\it J} }
_{\frac{\sqrt{2}}{3+\delta}} \left[ \frac{2\sqrt{\epsilon}}{3+\delta}
a^{(3+\delta)/2} \right] \nonumber\\&+& C_2 \mbox{ \large{\it Y} }
_{\frac{\sqrt{2}}{3+\delta}} \left[ \frac{2\sqrt{\epsilon}}{3+\delta}
a^{(3+\delta)/2} \right]~,
\eeq
where $J$ and $Y$ are Bessel functions of the first kind and second kind,
respectively, $C_1$ and $C_2$ are integration constants; for clarity the
$\phi$ dependence has been suppressed. We can thus evaluate approximations
to the derivatives:
\beq
\label{eq:dpsi}
 \partial_a \psi\left( a\right) &\approx& \left( 1 - \sqrt{\epsilon}Z_1\left(a\right)
 a^{(3+\delta)/2}\right)a^{-1}\psi\left(a\right)~, \nonumber\\
\label{eq:d2psi}
\partial^2_a\psi\left(a\right) &\approx& \left( Z_1\left( a\right)
a^{(3+\delta)/2}-\sqrt{\epsilon}a^{3+\delta} \right) \sqrt{\epsilon}a^{-2}
\psi\left(a\right)~,\nonumber \\
\label{eq:d3psi}
\partial^3_a\psi\left(a\right) &\approx& \left( \epsilon Z_1 a^{(9+3\delta)/2}
-3Z_1 a^{(3+\delta)/2} \right)
\sqrt{\epsilon}a^{-3}\psi\left( a\right)
\nonumber\\
&&-\sqrt{\epsilon}\left( 1+\delta\right)a^{3+\delta}
\sqrt{\epsilon}a^{-3}\psi\left( a\right)~, \\
&{\rm etc.}& \nonumber
\eeq
where
\beq
 Z_1\left( a\right) \equiv  C_1\mbox{ \large{\it J}}_{\frac{
\sqrt{2}}{3+\delta}+1}  \left[ \frac{2\sqrt{\epsilon}}{3+\delta}
a^{(3+\delta)/2}\right]\Bigl[ \psi\left(a\right) \Big]^{-1} \nonumber\\
+ C_2\mbox{ \large{\it Y}}_{\frac{\sqrt{2}}{3+\delta}
+1} \left[ \frac{2\sqrt{\epsilon}}{3+\delta} a^{(3+\delta)/2}\right]
\Bigl[ \psi\left(a\right) \Bigr]^{-1}~.
\eeq
Taking the large argument expansion of the Bessel functions, which is the 
large $a$ limit for $\delta > -3$, one obtains
\beq
 \lim_{a\rightarrow \infty} Z_1\left(a\right) &=& 
\frac{C_1\sin\left(A\left( a\right)\right) - C_2\cos\left(A\left( a
\right)\right)}{C_1\cos\left(A\left( a\right)\right)+C_2\sin\left(
A\left( a\right)\right)}~,\\
&=& \lim_{a\rightarrow \infty}\left[ \frac{\psi^*\left(
a\right)}{\psi\left( a\right)}\right]~,
\eeq
where $\psi^*\left(a\right)$ is the solution corresponding to choosing
the integration factors $C_1\rightarrow - C_2$ and $C_2\rightarrow
C_1$ compared to the solution given in Eq.~(\ref{eq:sol}). This
approximation is valid only when the correction terms are still small
compared to the matter component, however this method will give the
correct effective {\it initial} behaviour of QG corrections. In
particular, the corrections become non-negligible for scales at which
the corrections are of the order of $a_0^{-1}$ and the approximation
breaks down when they are of the order of $\epsilon\left( \phi \right)
a^\delta$. It is clear that for $\delta > -1$ the derivative terms
grow, in agreement with Ref.~\cite{Nelson:2007um}.  In addition, for
$\delta > -1$, the terms containing the highest powers of $\epsilon$
grow fastest and hence are always dominant, implying
\beq
\label{eq:dpsi2}
\partial_a\psi\left(a\right) &\approx& -\sqrt{\epsilon}\lim_{a\rightarrow \infty}
\left[ \frac{\psi^*\left( a\right)}{\psi\left(a\right)}\right] a^{(1+\delta)/2}
\psi\left(a\right)~,\nonumber \\
\label{eq:d2psi2}
\partial^2_a\psi\left( a\right) &\approx& -\epsilon a^{1+\delta}\psi\left(
a\right)~, \nonumber \\
\label{eq:d3psi2}
\partial^3_a\psi\left(a\right) &\approx& \epsilon^{3/2} \lim_{a\rightarrow \infty}
\left[ \frac{\psi^*\left( a\right)}{\psi\left( a\right)}\right]a^{3(1+\delta)/2}
\psi\left(a\right)~, 
\nonumber\\ 
\label{eq:d4psi2}
\partial^4_a\psi\left(a\right) &\approx& \epsilon^2 a^{2(1+\delta)}\psi\left(a
\right)~, \\
&{\rm etc.}& \nonumber 
\eeq
This allows us to approximate the initial effect of large scale break
down of classical physics (breakdown in {\it
preclassicality}~\cite{Bojowald:2002xz}) as a second matter
component. In general the $\beta^{th}$ derivative is proportional to
$a^{\beta(1+\delta)/2}$, thus correction terms like $ a^\alpha
\partial_a^\beta \psi\left(a\right)$ can be approximated by additional
terms in Eq.~(\ref{eq:WdW1}), that scale like $a^{
\alpha+\beta(1+\delta)/2}$. It is then possible to consider these
correction terms as being produced from an effective Hamiltonian.
A useful pedagogical example is to consider classical matter
terms (although the method is, of course, general), where the
effective Hamiltonian that mimics the correction terms is
\be
\label{eq:ham_cor}
 {\cal H}_{\rm cor} \propto a^{\alpha + \beta(1+\delta)/2}~.
\ee
An example of how these approximate solutions compare to the exact
case is given in Fig.~(\ref{fig1}).

We have shown how hypothetical correction terms to the WDW equation can mimic the
behaviour of additional matter sources.  In particular, such
corrections can be used to produce an effective cosmological constant.
For example, consider ${\cal H}_\phi \propto a^0$, i.e., a matter
dominated universe (see Table~\ref{table}); $\delta=0$. If we want the
quantum corrections to mimic a vacuum energy then we need ${\cal
H}_{\rm cor} \propto a^3 $, implying $\alpha+\beta/2=3$. Thus,
correction terms like $a_0a^3, ~a_0a^2 {\rm d}^2 \psi/{\rm d}a^2,
~a_0a{\rm d^4}\psi(a)/{\rm d}a^4, ~a_0 {\rm d}^6 \psi(a)/{\rm d}a^6,
{\rm etc.}$, with $a_0 < 0$, all resemble a vacuum energy in the presence of a matter
dominated universe (see Fig.~(\ref{fig1})). In the presence of other
types of matter $\delta$ changes and so the form of the correction
terms that give vacuum energy like behaviour are different. In
particular, for a universe dominated by radiation, we have $\delta=-1$
and the only correction term that can mimic the behaviour of dark
energy is $a_0 a^3$.  Considering a universe dominated by a field that scales
like ${\cal H}_\phi \propto a^2$, i.e., $\delta=2$, then a correction
term like $a_0a^{-3}{\rm d}^4 \psi(a)/{\rm d} a^4$ mimics a (negative)
cosmological constant like term. This is precisely the dominant correction
from LQC.

Up to now we have only discussed how the correction terms scale with
$a$, however there is clearly also a dependence on the constant part
of the matter Hamiltonian, $\rho_0$ (through
$\epsilon\left(\phi\right)$). In principle this differentiates these
correction terms from other models of dark energy (cosmological
constant, quintessence, etc.), as does the fact that as these
correction terms begin to dominate the WDW equation this approximation
will break down and the behaviour of the wave-function will be
drastically altered.

%%%%%%%%%%%%%%%%%%%%%%%%%%%%%%%%%%%%%%%%%%%%%%%%%%%%%%%%%%%%%%%%%%%%%%%%%%%%%%
%%%%%%%%%%%%%%%%%%%%%%%%%%%%%%%%%%%%%%%%%%%%%%%%%%%%%%%%%%%%%%%%%%%%%%%%%%%%%%

In addition to corrections that mimic the behaviour of dark energy, it
is also possible to find correction terms that have a dark matter like
form, for which Eq.~(\ref{eq:ham_cor}) is
\be
 {\cal H}_{\rm cor} \propto a^{\alpha+\beta(1+\delta)/2} \propto a^0~.
\ee
For example, in a matter dominated universe (i.e., $\delta = 0$),
\be
a_0a^{-1/2}\frac{{\rm d}\psi\left(a\right)}{{\rm d}a}, \ \ \ 
a_0a^{-1}\frac{{\rm d}^2\psi\left(a\right)}{{\rm d}a^2}, \ \ \
a_0a^{-3/2}\frac{{\rm d}^3\psi\left(a\right)}{{\rm d}a^3}, \ \ \ {\rm etc.}
\ee
all produce additional matter like terms. Notice that, unlike the dark energy
case, these corrections do not scale faster than original matter component
and so will never dominate Eq.~(\ref{eq:WdW2}). In addition, 
these correction matter terms will be closely related to the physical
matter Hamiltonian degrees of freedom. For example,
\be
 a_0a^{-2}\frac{{\rm d}^4\psi\left(a\right)}{{\rm d}a^4} \approx
 a_0\epsilon\left(\phi\right)^2\psi\left(a\right)~,
\ee
which amount to replacing $\epsilon\left(\phi\right)\rightarrow
\tilde{\epsilon} \left(\phi\right)=\epsilon\left(\phi\right)
+a_0\epsilon\left(\phi\right)^2$ in Eq.~(\ref{eq:WdW2}). Within the
mini-superspace model used here it is only possible to explore
homogeneous, isotropic solutions and so we cannot say that these dark
matter like correction terms would produce the necessary behaviour to
explain structure formation, galaxy and cluster dynamics, etc. Another
difficulty with this type of correction comes from the fact that the
energy density of dark matter is approximately five times that of
standard matter. If correction terms were to produce dark matter in
the presence of a matter dominated universe, $a_0$ would have to be
fine tuned to ensure that $a_0 \ll \epsilon\left(\phi\right)$ (the
approximation used here) and $a_0\epsilon\left(\phi\right)^{\beta/2-1}
\approx 5$. Similar dark matter like correction terms can be produced
for a radiation, or vacuum energy, dominated universe; if these
corrections arose due to the the presence of a vacuum energy, one may
explain the coincidence problem, i.e., that the dark matter degrees of
freedom would be dictated by those of the dark energy, however again
significant tuning would probably be required.

%%%%%%%%%%%%%%%%%%%%%%%%%%%%%%%%%%%%%%%%%%%%%%%%%%%%%%%%%%%%%%%

LQG is a background independent, non-perturbative method of quantising
gravity. Reducing the symmetries of the theory to a specific
cosmological model makes the theory tractable and ensures a large
scale continuum limit. The theory is based on holonomies of the
standard Ashtekar variables~\cite{Ashtekar:2003hd}, namely the triad
and connections. In an isotropic model these can be parameterised by
single variables, $\tilde{p}$ and $\tilde{c}$ respectively, which in
terms of standard cosmological variables are $ |\tilde{p}|=a^2,
\tilde{c}=k+\gamma \dot{a},$ ($\gamma$ is an ambiguity parameter known
as the Barbero-Immirzi parameter). Define $p=\tilde{p} V_0^{2/3}$ and
$c=\tilde{c}V_0^{1/3}$, where $V_0$ is the volume of a fiducial
cell~\cite{Vandersloot PhD}, related via the classical identity, $\{
c,p\} = 8\pi l_{\rm Pl}^2 \gamma/3$.  By analogue with the full LQG
theory, $c$ is quantised via its holonomies, $h=\exp (i\mu_0 c/2)$,
where $\mu_0$ is an arbitrary real number. Then~\cite{Vandersloot
PhD}, $\hat{p} |\mu\rangle =|p||\mu\rangle \equiv \frac{4\pi l_{\rm
Pl}^2\gamma|\mu|}{3}| \mu\rangle,$ and $ \widehat{ e^{\mu_0 c/2}}
|\mu\rangle =e^ {\mu_0 \frac{{\rm d}}{{\rm d}\mu } }|\mu
\rangle=|\mu+\mu_0\rangle.$

The action of the gravitational part of the Hamiltonian constraint on
the basis states $|\mu \rangle$ reads~\cite{Vandersloot:2005kh},
\beq\label{eq:ham_dis}
\hat{\mathcal H}_{\rm g} | \mu \rangle = \frac{3}{256\pi^2l_{\rm Pl}^4 \gamma^3
 \mu_0^3} \Bigl\{ \bigl[ S(\mu)+S(\mu+4\mu_0)\bigr]| \mu+4\mu_0
\rangle \nonumber\\-4S(\mu ) | \mu\rangle +\bigl[ S(\mu)+S(\mu-4\mu_0)\bigr]| 
\mu-4\mu_0 \rangle \Bigr\}~,
\eeq
where 
\be
S(\mu )= \left( 4\pi l_{\rm Pl}^2 \gamma /3\right)^{3/2} 
\Big| | \mu+\mu_0 |^{3/2} -|\mu-\mu_0|^{3/2} \Big|~.
\ee
The Hamiltonian constraint is $\left(\hat{\cal H}_{\rm g} + \hat{\cal H}_\phi
\right)|\psi\rangle = 0$, where $|\psi\rangle =\sum_\mu \psi_\mu|\mu\rangle$.
Taking the continuum limit $\psi_\mu\rightarrow \psi\left(\mu\right)$ and
using the expansion~\cite{Nelson:2007wj}
\beq
 | \mu+\alpha\mu_0|^{3/2} - |\mu+\beta\mu_0|^{3/2} = \mu^{3/2} \Biggl[ 
\frac{3\mu_0}{2\mu} \left(\alpha-\beta\right)\nonumber\\
+\frac{3\mu_0^2}{8\mu^2}\left( \alpha^2-\beta^2\right) -\frac{\mu_0^3}{16\mu^3}
\left( \alpha^3-\beta^3 \right) +\cdots \Biggr]~,
\eeq
we can expand Eq.~(\ref{eq:ham_dis}).

Changing variables, using $a^2=4\pi l_{\rm Pl}^2\gamma|\mu|/3$, we
obtain 
\beq \label{eq:LQC_H}
&&{\cal H}_{\rm g} \psi\left(a\right) = \frac{2\pi l_{\rm
Pl}^2}{3} \Biggl( \frac{1}{a}\frac{{\rm d}}{{\rm d} a}
\Biggl[\frac{1}{a}\frac{{\rm d}}{{\rm
d}a}\bigl[a\psi\left(a\right)\bigr]\Biggr] + \frac{{\rm d}}{ {\rm d}
a} \Biggl[ \frac{1}{a}\frac{{\rm d}\psi\left(a\right)}{{\rm
d}a}\Biggr] \Biggr) \nonumber\\ 
&&+ \frac{2\pi l_{\rm Pl}^2 a_0}{9}
\Biggl( \frac{1}{a^3}\frac{{\rm d}^4\psi\left(a \right)}{{\rm d} a^4}
-\frac{4}{a^4}\frac{{\rm d}^3 \psi\left(a\right)}{{\rm d}a^3}
+\frac{47}{8a^5}\frac{{\rm d}^2 \psi\left(a\right)}{{\rm
d}a^2}\nonumber\\ 
&&+\frac{1}{2a^6}\frac{{\rm d}\psi\left(a
\right)}{{\rm d}a} -\frac{135}{16a^7} \psi\left(a\right) \Biggr)~,
\eeq
where $a_0=16\pi^2l_{\rm Pl}^4\gamma^2\mu_0^2/9$. The full Hamiltonian
constraint is easily written in the same form as
Eq.~(\ref{eq:WdW2}). As we have shown, for $\delta>-1$ the higher
derivative terms scale faster. Since in this case the lower derivative
terms are further suppressed by powers of $a$, to a good approximation
the effective large scale equation to solve is
\beq \frac{1}{2}\left(\frac{1}{a}\frac{{\rm d}}{{\rm d} a}
\Biggl[\frac{1}{a}\frac{{\rm d}}{{\rm
d}a}\bigl[a\psi\left(a\right)\bigr]\Biggr] + \frac{{\rm d}}{ {\rm d}
a} \Biggl[ \frac{1}{a}\frac{{\rm d}\psi\left(a\right)}{{\rm
d}a}\Biggr]\right)\nonumber\\ +\frac{a_0}{3}
\epsilon\left(\phi\right)^2a^{2\delta-1}\psi\left(a\right)
+\epsilon\left(\phi\right)a^\delta\psi\left(a\right)\approx 0~,
\eeq
where $3{\cal H}_\phi/(2 \pi l_{\rm Pl}^2) = \epsilon\left(\phi
\right) a^\delta$ and we used Eq.~(\ref{eq:d4psi2}d) to approximate
the fourth order derivative term in Eq.~(\ref{eq:LQC_H}) by term that
resembles an additional matter component.

LQC is a concrete example of the general methods for approximating the
QG corrections. For a universe dominated by a (classical) matter content with
$\delta=0$, the leading correction term acts as an effective radiation
field, ${\cal H}_{\rm cor} = a_0 \epsilon\left(\phi\right)^2a^{-1}/3$,
i.e., the correction term mimics radiation. For a radiation dominated
universe with $\delta = -1$, the correction term acts as ${\cal
H}_{\rm cor} = a_0 \epsilon \left(\phi \right)^2a^{-3}$. For $\delta =
3$, i.e., a universe dominated by a vacuum energy, such as during
inflation, the correction term acts like ${\cal H}_{\rm cor} = a_0
\epsilon \left(\phi\right)^2a^5$. Notice that in this case the
corrections scale faster than the existing matter component, which
motivated the modelling of {\it lattice refinement} ($\mu_0\rightarrow
\tilde{\mu}\left(\mu\right)$ is no longer a constant) in loop quantum
cosmology~\cite{Nelson:2007um,Bojowald:2002xz,Ashtekar:2006uz}.
For a universe dominated by a source with $\delta=2$, the dominant
correction term acts as ${\cal H}_{\rm cor } =  a_0\epsilon\left( \phi
\right)^2a^3$, i.e. it mimics a negative cosmological constant (i.e. an
accelerated {\it contraction}), whilst $\delta = 1/2$ leads to the
correction term mimicking dark matter.

%%%%%%%%%%%%%%%%%%%%%%%%%%%%%%%%%%%%%%%%%%%%%%%%%%%%%%%%%%%%%%%%%%%%%%%%%%%%%%
%%%%%%%%%%%%%%%%%%%%%%%%%%%%%%%%%%%%%%%%%%%%%%%%%%%%%%%%%%%%%%%%%%%%%%%%%%%%%%

We have shown how generic QG corrections to the WDW equation mimic the
behaviour of additional matter sources, at least whilst they are
smaller than the existing matter components. This simple procedure can
be used to examine, to first order, the effect of any QG correction
under consideration. Here, as a specific example, we used the first
order corrections from loop quantum cosmology to examine what the
breakdown of large scale classical physics looks like in the absence
of lattice refinement. The fact that quantum corrections introduce extra
matter components opens up the possibility that they may be able to
explain the current cosmological acceleration and possibly dark matter.
That these correction components become significant only at specific scales
is encouraging, as this may provide an explanation why classical general
relativity is valid on sub-galactic scales, but requires the input of
addition matter on super-galactic scales. However it remains to be seen
if it is possible for such corrections to meet observational constraints.

This work can be used in two complementary ways: top-down and
bottom-up. The former is to apply this method to any fundamental
theory that can produce a WDW like equation on large scales and so
characterise the new, phenomenological effects of the theory. The
latter would estimate the types of QG corrections that would be
necessary to produce certain desirable effects (e.g., dark energy) at
particular scales.  The case of loop quantum cosmology illustrates how
both approaches can be used to mutual benefit. The underlying theory
was used to calculate the form of the QG corrections to the WDW
equation. The phenomenological consequence (eventual domination of the
correction terms and hence a break down in large scale classical
physics) highlighted the importance of modelling lattice refinement
within cosmological models.

Considering the phenomenological consequences of a theory is the first
step to testing it against experimental and observational data. We
have shown here that for QG this principle can be invaluable in
guiding our search for the full theory.

\acknowledgements This work is partially supported by the European
Union through the Marie Curie Research and Training Network
\emph{UniverseNet} (MRTN-CT-2006-035863).
%%%%%%%%%%%%%%%%%%%%%%%%%%%%%%%%%%%%%%%%%%%%%%%%%%%%%%%%%%%%%%%%%%%%%%%%%%%%%%%

\end{document}